# Enhancing brain age estimation with structural MRI and synthesized cerebral blood volume maps

Jordan Jomsky,[1] Zongyu Li,[1] Kay C. Igwe,[1] Yiren Zhang,[1] Max Lashley,[2] Tal Nuriel,[3,4] Andrew Laine,[1] Scott A. Small,[3,5,6,7,8] Jia Guo[2,6] and for the Frontotemporal Lobar Degeneration Neuroimaging Initiative and for the Alzheimer's Disease Neuroimaging Initiative[†]

† Data used in preparation of this article were obtained from the Alzheimer's Disease Neuroimaging Initiative (ADNI) database (adni.loni.usc.edu). As such, the investigators within the ADNI contributed to the design and implementation of ADNI and/or provided data but did not participate in analysis or writing of this report. A complete listing of ADNI investigators can be found at: http://adni.loni.usc.edu/wp-content/uploads/how_to_apply/ADNI_Acknowledgement_List.pdf

Brain age gap estimation (BrainAGE) is a promising imaging-derived biomarker of neurobiological ageing and disease risk, yet current approaches rely predominantly on T1-weighted structural MRI, overlooking functional vascular changes that may precede tissue damage and cognitive decline. Deep learning-derived cerebral blood volume (DeepCBV) maps, synthesized from non-contrast MRI, offer a scalable alternative to contrast-enhanced perfusion imaging by capturing vascular information relevant to early neurodegeneration. We developed a multimodal BrainAGE framework that combines predictions from two separate three-dimensional convolutional neural networks: one trained only on structural MRI scans and another trained only on DeepCBV maps generated by a pre-trained three-dimensional patch-based deep learning model. Each model was trained and validated on 2851 scans (1507 females) from 13 open-source datasets and was evaluated for concordance with mild cognitive impairment (MCI) and Alzheimer's disease (AD) using 1233 subjects. The combined model achieved the most accurate brain age gap for cognitively normal (CN) controls, with a mean absolute error of 3.95 years ($R^2 = 0.943$), outperforming models trained on MRI (mean absolute error = 4.10) or DeepCBV alone (mean absolute error = 4.49). Saliency maps revealed complementary modality contributions: MRI emphasized white matter and cortical atrophy, while DeepCBV highlighted vascular-rich and periventricular regions implicated in hypoperfusion and early cerebrovascular dysfunction, consistent with known patterns of normal ageing. Next, we observed that BrainAGE increased stepwise across diagnostic strata (CN < MCI < AD) and correlated with cognitive impairment (Clinical Dementia Rating Sum of Boxes $\rho = 0.403$; Mini-Mental State Examination $\rho = -0.310$). DeepCBV-based BrainAGE showed a particularly strong separation between stable versus progressive MCI (Mann–Whitney $U = 2.177 \times 10^4$, $P = 4.43 \times 10^{-8}$), suggesting sensitivity to prodromal vascular changes that precede overt atrophy. Integrating structural MRI with deep learning-derived vascular measures substantially enhances BrainAGE estimation and improves sensitivity to MCI and Alzheimer's disease progression, supporting its potential role in risk stratification, early detection and monitoring of therapeutic response. By enabling a functional-like assessment from routine MRI, this approach lowers barriers to multimodal evaluation and provides a clinically actionable biomarker for large-scale ageing and dementia studies.

1 Department of Biomedical Engineering, Columbia University, New York, NY 10027-7041, USA
2 Zuckerman Institute, Columbia University, New York, NY 10027-7041, USA
3 Taub Institute for Research on Alzheimer's Disease and the Aging Brain, Vagelos College of Physicians and Surgeons, Columbia University, New York, NY, USA
4 Department of Pathology and Cell Biology, Columbia University, New York, NY, USA
5 Department of Neurology, Columbia University, New York, NY 10027, USA
6 Department of Psychiatry, Columbia University, New York, NY 10027, USA
7 Department of Radiology, Columbia University, New York, NY, USA
8 New York State Psychiatric Institute, New York, NY, USA

Received December 10, 2025. Revised June 04, 2026. Accepted June 25, 2026. Advance access publication July 22, 2026

Correspondence to: Jia Guo, PhD
Zuckerman Institute and Department of Psychiatry, Columbia University in the City of New York. 3227 Broadway, New York, NY 10027, USA
E-mail: jg3400@columbia.edu

**Keywords:** magnetic resonance imaging; brain age; deep learning; cerebral blood volume

## Graphical Abstract

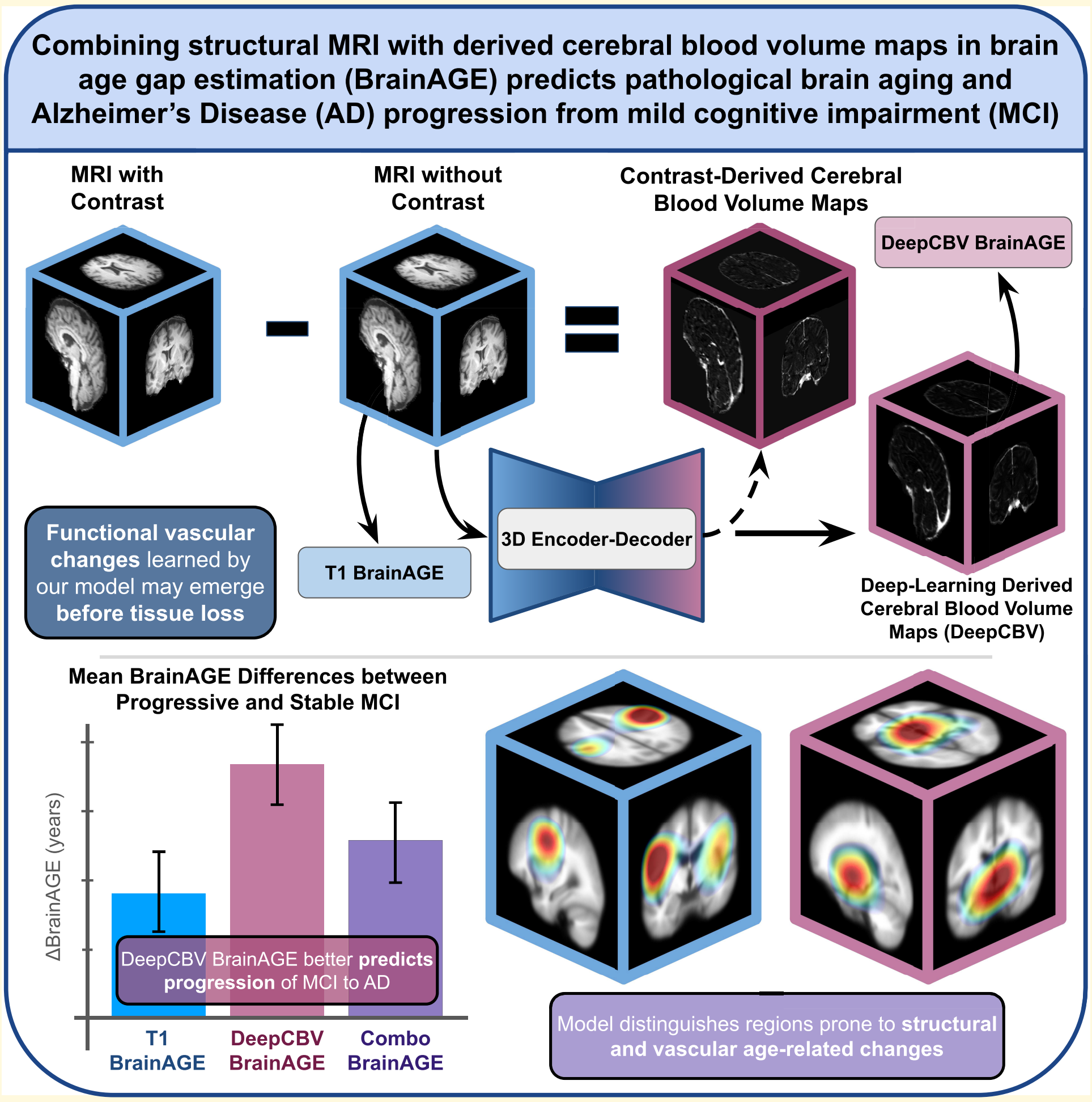

# Introduction

Ageing is a universal heterogeneous biological process that contributes to an increased risk of developing neurodegenerative diseases, such as Alzheimer's disease (AD).[1-3] This inherent variability in biological ageing, independent of chronological age, may be an indicator of pathological ageing, reflecting an acceleration of biological mechanisms, such as inflammation and aberrant protein accumulation, also present in healthy ageing.[4] In the brain, pathogenic mechanisms of biological ageing manifest as a collection of structural and functional changes, including cortical and subcortical atrophy,[5] white matter hyperintensity (WMH) accumulation,[6-8] widening perivascular spaces,[9,10] brain perfusion[11,12] and functional connectivity dynamics,[13] all of which are detectable through MRI. These non-invasive neuroimaging tools are instrumental in quantifying biological age, a measure of the deviation from chronological age that reflects pathological changes. This deviation from a chronological age is known as brain age gap estimation (BrainAGE), serving as a crucial biomarker for aberrant ageing.[14]

Brain age gap estimation (BrainAGE) has emerged as a promising neuroimaging-based biomarker for estimating a divergence from normative brain ageing, potentially revealing underlying pathophysiological changes associated with neurodegeneration. T1-weighted (T1w) MRI has become a primary non-invasive imaging modality for assessing BrainAGE due to its utility in capturing age-related anatomical changes, its wide availability across large open-source neuroimaging databases, and its clinical utility. Various studies have demonstrated that machine and deep learning models trained solely on T1w images can accurately predict chronological age from brain scans of healthy individuals.[15-20] In fact, Feng *et al.*[16] were able to accurately predict chronological age (mean absolute error = 4.06, $R^2$ = 0.941) using a Visual Geometry Group-based deep learning model and the T1w brain image as input. While numerous recent BrainAGE models have reported lower mean absolute error using a variety of architectures and large-scale datasets, including transformer-based and minimally preprocessed approaches, direct comparisons across studies are complicated by differences in cohort composition, preprocessing pipelines and evaluation strategies [19-25] We reference Feng *et al.*[16] primarily because of the use of a similar Visual Geometry Group-based architecture and similar training data, which permits a more direct comparison. Importantly, the primary contribution of the present work is not state-of-the-art age prediction accuracy per se, but the demonstration that a learned vascular representation derived from non-contrast structural MRI provides complementary information that improves BrainAGE estimation and clinical discrimination.

Although T1w-based BrainAGE models are effective in predicting chronological age, this approach may fail to account for underlying biological processes that reflect brain function. Further, recent studies have demonstrated that vascular changes and cerebral blood flow, as measured by functional data modalities, provide earlier and more accurate biomarkers of AD onset.[26,27] Combining structural and functional imaging modalities to determine BrainAGE may offer a more comprehensive clinical evaluation for ageing and age-related neurodegenerative diseases and provide insight into the divergence of brain age prediction from chronological age.

Cerebral blood volume functional MRI, which utilizes paired T1-weighted imaging acquired with and without gadolinium-based contrast agents (GBCAs), produces functional CBV maps that measure regional metabolism, providing a quantitative correlate of metabolic dysfunction. Further, the use of gadolinium-based contrast agents provides a measure of the blood–brain barrier (BBB) integrity through observation of contrast agent kinetics, all of which serve as critical measures for elucidating pathological ageing. However, the use of gadolinium-based contrast agents is invasive, requiring an injection, and is not feasible for large longitudinal studies. Additionally, CBV-fMRI is not as accessible in large-scale clinical datasets as T1w MRI.

We introduce Deep learning-derived Cerebral Blood Volume (DeepCBV) maps, a non-contrast surrogate for CBV-fMRI generated directly from standard T1-weighted MRI. This approach allows for the generation of high-resolution CBV maps without the use of contrast agents, making it suitable for large-scale and retrospective neuroimaging studies. We hypothesized that combining T1w MRI and DeepCBV maps into a single model may serve as a more accurate estimator of brain age and neurodegenerative risk, capturing both structural integrity and functional vascular changes that are critical in ageing and Alzheimer's disease progression.

Our study had two main objectives for assessing the efficacy of our combined T1w and DeepCBV approach. First, we determined whether training a model on both a structural T1w image and a deep learning-derived CBV map could reveal a more robust BrainAGE metric than either modality alone. Second, we examined whether our combined model could identify established Alzheimer's disease-specific relationships, which would demonstrate our model's clinical relevance.

# Materials and methods

This study did not require ethical approval as it relied on publicly available anonymized MRI scans that were originally collected as part of studies already approved by the requisite standards.

## Data sources

In this study, we leveraged neuroimaging and clinical data from 13 publicly available datasets: Alzheimer's Disease Neuroimaging Initiative (ADNI), Brain Genomics Superstruct Project (BGSP),[28] Neuroimaging in Frontotemporal Dementia (NIFD), Southwest University Longitudinal Imaging Multimodal (SLIM) study,[29] Dallas Lifespan Brain Study (DLBS),[30] Southwest University Adult Lifespan Dataset (SALD),[31] Australian Imaging, Biomarkers & Lifestyle Study of Aging (AIBL),[32] Consortium for Reliability and Reproducibility (CoRR),[33] SchizConnect,[34] Information eXtraction from Images (IXI),[35] Open Access Series of Imaging Studies (OAS1 and OAS2)[36,37] and the Parkinson's Progression Markers Initiative (PPMI).[38] Data used in the preparation of this article were obtained from the ADNI database (adni.loni.usc.edu). The ADNI database was launched in 2003 as a public–private partnership, led by Principal Investigator Michael W. Weiner, MD. The primary goal of ADNI has been to test if MRI, positron emission tomography (PET), other biological markers and clinical and neuropsychological assessment can be combined to measure the progression of mild cognitive impairment (MCI) and early AD.

The entire dataset included 2851 T1w MRI images (1507 females). All images were registered to the MNI-152 template, a widely used standard brain reference derived from averaging 152 structural MRI scans from healthy adults, using an affine transformation.[39] A diverse set of data sources in this study were used to enhance the transferability and generalizability of our approach to different datasets. The distribution of patient ages across the datasets, in each study and the data split used to train and validate the model can be seen in Fig. 1, where each plot highlights the variability in age ranges across each study-specific dataset used to train our model.

To validate our model's AD risk assessment, we utilized data from the ADNI, which includes individuals clinically categorized as cognitively normal (CN), MCI and Alzheimer's disease. ADNI subjects included in clinical analyses were fully excluded from BrainAGE model training, validation and regression fitting. All ADNI disease-related results were obtained by applying pre-trained models to ADNI data in inference mode only, ensuring strict separation between model development and clinical validation. Within the MCI cohort, participants were further stratified into stable MCI (SMCI) and progressive MCI (PMCI) based on longitudinal diagnostic trajectories. Cognitive and clinical severity were quantified using three well-established measures: the Clinical Dementia

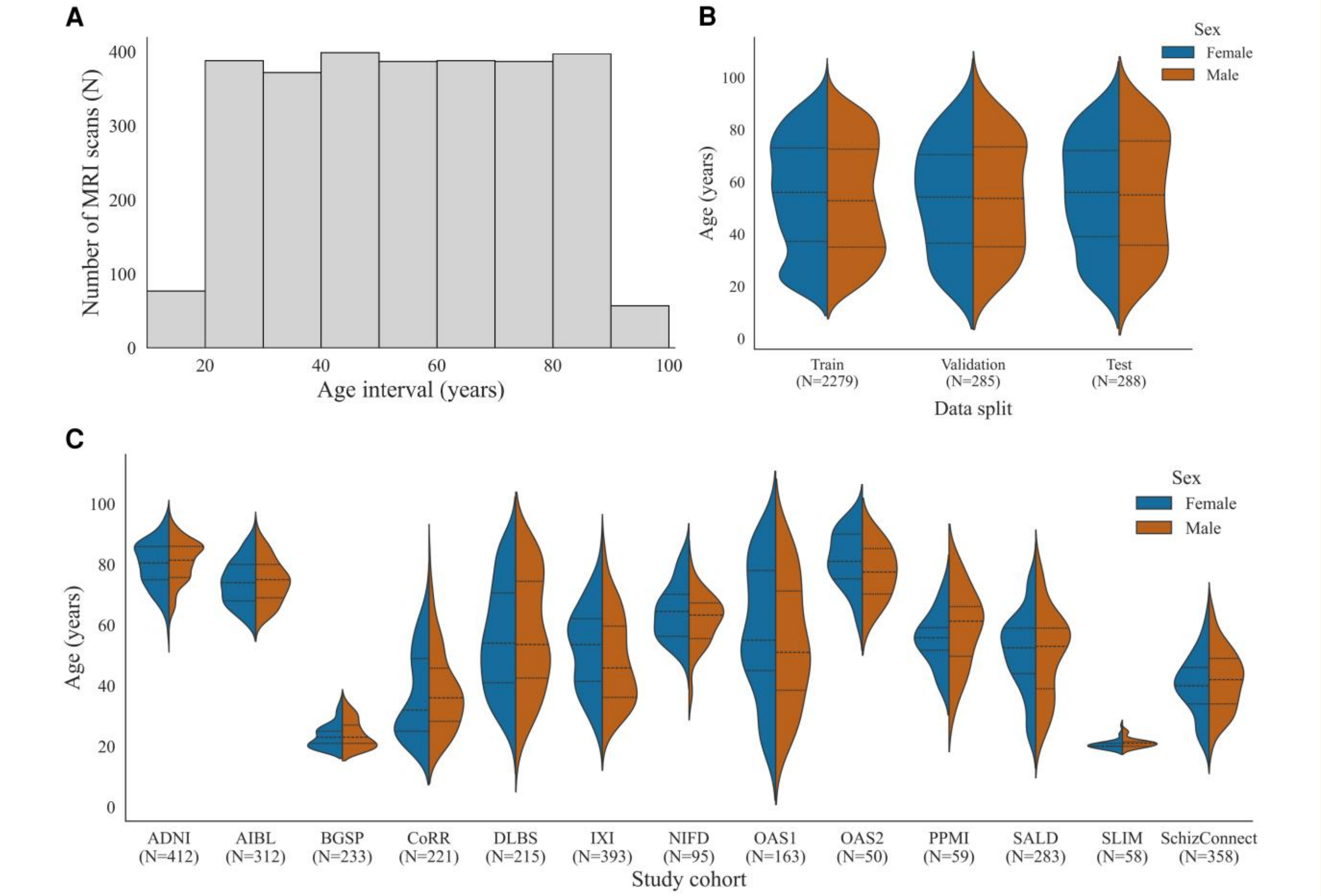

**Figure 1 Age distributions across datasets and model splits.** (**A**) Distribution of participant age across the full dataset (*N* = 2851 MRI scans), shown as counts per 10-year interval. (**B**) Distribution of age (years) by study cohort and sex across all included datasets. Each violin represents the distribution of individual MRI scans within that cohort (*N* per cohort shown on the *x*-axis). (**C**) Distribution of age (years) across the training, validation and test splits used for model development (total *N* = 2851; *N* per split shown on the *x*-axis). Each violin represents individual MRI scans.

Rating Sum of Boxes (CDRSB), which captures functional decline across multiple domains and is sensitive to the widespread consequences of brain atrophy[40]; the Mini-Mental State Examination (MMSE), a cognitive measure used for early detection of dementia[41]; and the Alzheimer's Disease Assessment Scale–Cognitive Subscale (ADAS13), a sensitive marker of cognitive decline across the spectrum from MCI to AD.[42]

## DeepCBV model

Conventional CBV-sensitive mapping in steady-state, gadolinium-based protocols relies on acquiring paired T1-weighted scans before and after contrast and then deriving a voxel-wise enhancement map that reflects contrast-related signal increase.[43] In practice, this enhancement proxy is obtained by intensity-matching the paired scans and computing the difference between the post-gadolinium and the pre-gadolinium image, which encodes the spatial pattern of gadolinium-related uptake and therefore contains vascular information not present in the non-contrast structural scan alone.[43]

Because most large-scale, open neuroimaging datasets used for brain age modelling do not include contrast-enhanced acquisitions, we generated DeepCBV maps using DeepContrast, a deep-learning pipeline developed to substitute gadolinium administration at inference time by predicting a voxel-wise contrast-uptake map from a single non-contrast T1w scan.[43-46] In this manuscript, DeepCBV refers to the DeepContrast-predicted voxel-wise relative cerebral blood volume (rCBV) map computed from non-contrast T1w alone and used as an additional CBV-sensitive vascular feature for downstream BrainAGE modelling.

### Training dataset and preprocessing for DeepContrast

As described by Liu *et al.*,[43] all DeepContrast model versions were trained on an in-house steady-state CBV-fMRI human dataset consisting of paired T1w and T1w-CE volumes acquired in the same session, with a train/validation/test split of 326/93/179 participants, respectively; receiver gain was held constant and offset set to zero so that voxel intensities remained physically comparable within each pair. Each T1w volume was intensity-compressed to the range [0, 1], and the corresponding T1w-CE volume was scaled by the same factor prior to computing the rCBV target by subtraction.[43] Loss evaluation was restricted to brain tissue using masks generated with FMRIB Software Library Brain Extraction Tool.[47]

### Supervision target

rCBV was defined as the voxel-wise gadolinium-induced signal enhancement normalized by the enhancement measured from pure blood. The scalar-matched pre-contrast T1-weighted (T1w) and contrast-enhanced T1-weighted (T1w-CE) images were first spatially aligned. A gadolinium uptake map was then generated by subtracting the T1w image from the T1w-CE image.[43] Because the two images were scalar-matched, this signal difference was assumed to be proportional to local gadolinium uptake.

To obtain rCBV, the gadolinium uptake map was normalized by the mean signal enhancement measured within a pure blood region of interest, such as a large vessel or venous sinus. The resulting dimensionless map represents blood-normalized gadolinium enhancement and was interpreted as a relative CBV measure.[43,44] Higher rCBV values indicate greater blood-volume-related gadolinium uptake relative to the blood reference signal.

### DeepContrast models

To ensure methodological completeness, we summarize the four DeepContrast model generations that underpin our DeepCBV generation pipeline. We provide a brief architectural walkthrough to contextualize the model used for generating voxel-wise rCBV maps as input for the BrainAGE model.

### 2D RA-UNet and 3D extension

The first working DeepContrast model version was constructed using a residual attention U-Net style encoder-decoder trained on coronal slices.[43] Motivated by the inherently 3D continuity of anatomy and vascular patterns we extended this model architecture to a volumetric residual-attention U-Net that processes 3D context directly.

### Transformer-enhanced U-Net

We next evaluated the Transformer U-Net hybrid TransUNet encoder representations to better capture long-range dependencies that can be difficult for purely convolutional receptive fields,[48] while maintaining an image-to-image regression head that outputs voxel-wise rCBV maps. This direction is supported by prior work showing that Transformer-based representations can improve robustness and cross-dataset generalization in brain MRI tasks.[49]

### Patch-based 3D Transformer

We implemented a patch-based 3D Transformer that predicts voxel-wise rCBV from overlapping subvolumes and reconstructs a full-volume map by averaging overlapping predictions. This design is inspired by Transformer U-Net hybrid approaches that leverage long-range Transformer context in brain MRI.[48,50] Each T1w scan was conformed to MNI template space and resampled to 1 mm isotropic resolution, then padded or cropped to $192 \times 192 \times 160$ voxels.[39] Overlapping $96 \times 96 \times 96$ patches were extracted with a sliding-window strategy with a stride of six voxels, each patch was independently mapped to predict a rCBV patch, and overlapping predictions were averaged voxel-wise to reduce blocking artefacts and improve spatial consistency.[50]

### DeepContrast training protocol

All models were optimized using stochastic gradient descent with a batch size of 4 and a robust adaptive regression loss. Evaluation was performed with the brain-extracted scans. DeepContrast synthesis fidelity was quantified by whole-brain, voxel-wise comparisons between predicted maps and paired ground-truth rCBV maps using peak signal-to-noise ratio (PSNR, decibels [dB]) and structural similarity index measure (SSIM).

### Full-cohort DeepCBV generation for BrainAGE

Following synthesis-fidelity benchmarking across model generations, the selected DeepContrast model was applied at inference time to the full multi-dataset cohort to generate DeepCBV maps from non-contrast T1w input alone, enabling CBV-sensitive vascular proxy feature extraction in datasets lacking contrast-enhanced scans.

## BrainAGE model

Our BrainAGE model was implemented using a VGG8 architecture in PyTorch.[51] Each block consisted of a 3D convolutional layer, followed by 3D batch normalization, a rectified linear unit (ReLU) activation layer and a 3D max pooling layer, with five repeated blocks making up the encoder portion of the model. The number of output channels in each convolutional layer was 16, 32, 64, 128 and 256. The final three fully connected layers map the flattened encoder output to 512, then 128 and finally one unit to produce the final brain age prediction.

A linear regression model was implemented in R (Version 4.4.1) to combine the predictions from each model. Each model's prediction was combined and used as a feature to predict the actual age of the patient. This combined approach integrates structural and functional age estimates, leveraging the unique strengths of each model. Additionally, we used sex as an additional covariate in our final model. Figure 2 showcases the full model pipeline from data collection to inference.

The dataset was first stratified based on age. Specifically, the age distribution was divided into deciles, which identified the 10th to the 90th percentiles of the age variable. The resulting quantiles, along with the minimum and maximum age values, were used to define the bin edges, and each scan was accordingly assigned to these age bins. The dataset was partitioned into ten separate stratified partitions based on age bins as well as the project origin of the scan. The dataset of 2851 scans was split in an 8:1:1 ratio into training, validation and test sets. The validation and test set were selected from the ten partitions with minimal Kullback–Leibler (KL) divergence compared to the dataset overall to ensure generalizable performance.

Preprocessing steps were previously described in Feng *et al.*,[16] and were applied without modification. Briefly, T1w MRI images from all cohorts were first intensity bias field corrected,[52] then brain extracted,[53] and finally registered to the standardized MNI-152 affine space.[39] All images were manually checked at each stage to ensure image quality, whole-brain coverage and the integrity of intermediate preprocessing outputs. We applied a modified intrascan min-max normalization procedure to each T1w image, where each voxel is divided by the mean of the top 1% voxel intensities in that scan. Images were excluded only if the original T1w scan had insufficient whole-brain coverage, image-obstructing artefacts were present or a poor-quality image after any preprocessing step. For this current study, no images were excluded from the dataset we used.

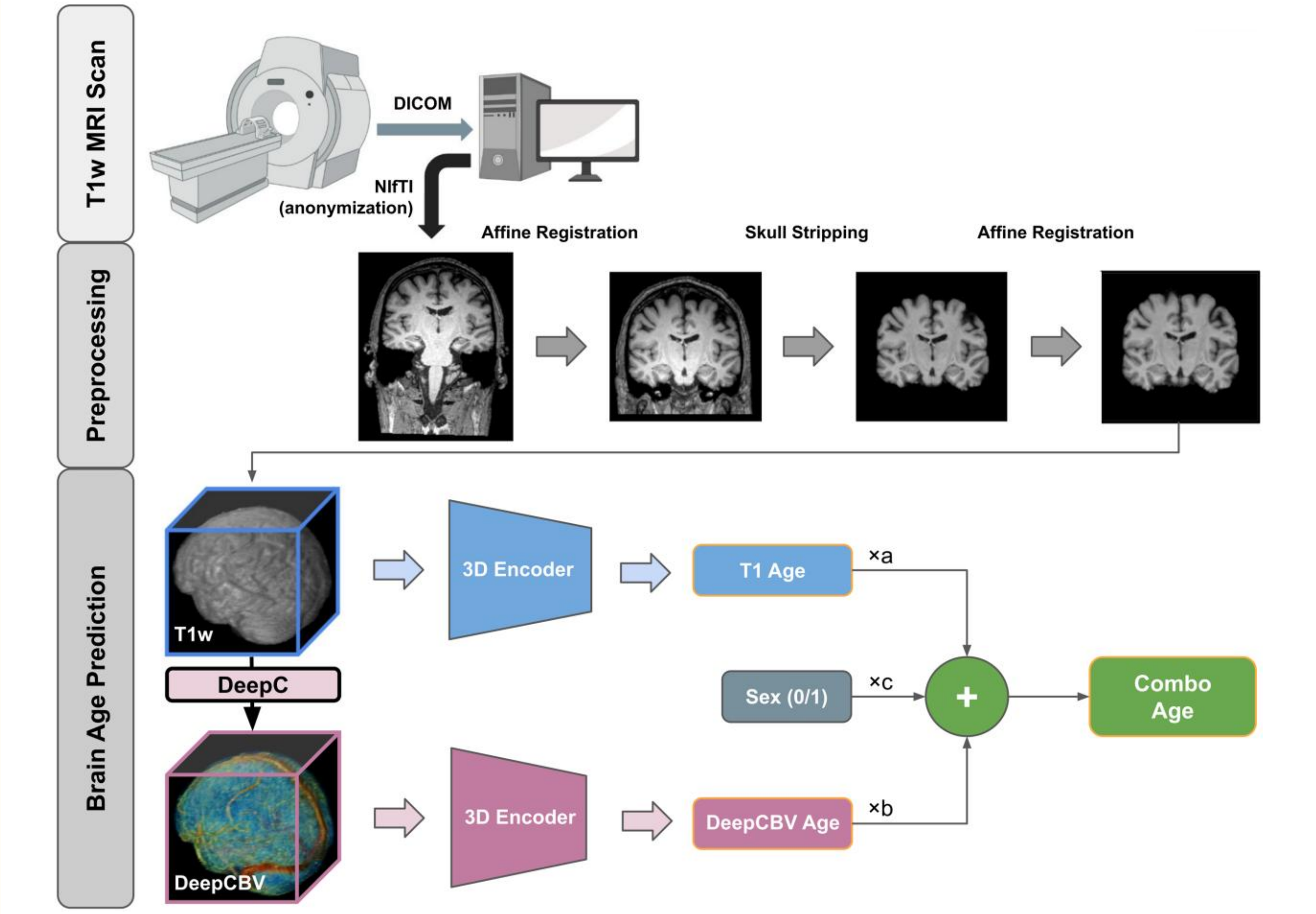

**Figure 2 Brain age estimation pipeline.** Overview of the multimodal framework used to estimate brain age from structural MRI. Raw T1-weighted (T1w) MRI scans are acquired in Digital Imaging and Communications in Medicine (DICOM) format and converted to Neuroimaging Informatics Technology Initiative (NIfTI) format following anonymization. Preprocessing includes affine registration to a standardized template space, skull stripping (removal of non-brain tissue), and final affine alignment. The preprocessed T1w volume is input into two parallel pathways. In the first pathway, the T1w scan is passed through a three-dimensional (3D) convolutional encoder to generate a predicted structural brain age ('T1 Age'). In the second pathway, a deep learning-based contrast synthesis model (DeepC) generates deep learning-derived cerebral blood volume maps (DeepCBV) from the same T1w scan. The DeepCBV volume is then processed through a separate 3D encoder to produce a predicted vascular-based brain age ('DeepCBV Age'). Predicted ages from the T1w and DeepCBV encoders are combined using a linear regression model. The symbols ×a and ×b denote learned regression coefficients applied to the T1 Age and DeepCBV Age predictions, respectively. Sex is included as an additional covariate (coded as 0/1) and multiplied by coefficient ×c. The green '+' node represents linear combination of these weighted inputs. The final output ('Combo Age') represents the multimodal brain age prediction integrating structural, vascular and sex-related information.

The Visual Geometry Group model for both T1w and DeepCBV scans was trained for 100 epochs with early stopping if the mean absolute error (MAE) did not improve after 10 epochs. An AdamW optimizer with a learning rate of $1 \times 10^{-4}$ and MAE loss was used to train the model with a batch size of 3.[54] All models were trained on the NVIDIA RTX A6000 GPU. After training the modality-specific deep learning models, predictions were generated separately for the training, validation and held-out test sets. Predictions from the training and validation sets were used to fit a linear regression model combining T1w- and DeepCBV-derived age estimates. The fitted regression parameters were then applied to predictions from the held-out test set to generate final BrainAGE estimates for performance evaluation. All models and weights can be obtained from the following GitHub repository: https://github.com/jzjomsky/DeepCBV-BrainAGE.

## Statistical analysis

Linear regression analysis was performed using the chronological age of the patient as the dependent variable and the predicted age and the sex of the patient from each model as features with male coded as 1 and female coded as 0. We computed these results along with their residuals and coefficient of determination ($R^2$). In order to determine if the multimodal models showed statistically significant improvement over the unimodal models, we also performed a bootstrap analysis of the residual sum of squares (RSS).

To examine the validity of our BrainAGE estimates with known biomarkers of AD, we computed the Spearman correlation coefficients (ρ) between BrainAGE values and each cognitive score. We then evaluated whether BrainAGE estimates differed across clinical stages of AD progression. To assess this, a Kruskal–Wallis test was used to detect whether there was a significant difference in brain age gaps between CN, MCI and AD BrainAGE estimates from each model. After ensuring statistically significant results from these initial tests, pairwise Mann–Whitney tests were performed for each combination of AD diagnosis strata within each model. Similar Mann–Whitney tests were used to determine statistical significance in brain age gaps for each model in the SMCI and PMCI groups. Bonferroni correction was applied to correct for multiple comparisons ($\alpha = 0.05$). Additionally, we assessed group-level differences in BrainAGE using Cohen's d effect sizes. Together, these analyses were used to evaluate both the continuous associations between BrainAGE and cognitive decline as well as the discriminative ability of BrainAGE to differentiate between clinical stages of disease progression.

# Results

## DeepCBV synthesis across DeepContrast model generations

We evaluated DeepCBV synthesis fidelity by comparing DeepContrast-predicted maps against the ground-truth voxel-wise rCBV targets derived from paired pre-contrast and post-contrast T1w acquisitions. As summarized in Table 1, synthesis fidelity improved across successive DeepContrast model generations, consistent with architectural changes that progressively increase the ability to model long-range context and global structure beyond local convolutional neighbourhoods.[48,50]

The patch-based 3D Transformer model achieved the strongest overall fidelity by PSNR and SSIM and better recovered vascular patterns aligned with the rCBV target in representative slices shown in Fig. 3. The stronger agreement between prediction and ground truth, relative to the raw T1w comparison, supports that the predicted DeepCBV maps encode contrast-related structure not directly present in the non-contrast image.[43] Based on these results, our patch-based 3D TransUNet DeepContrast model was selected as the DeepCBV generator and used to produce DeepCBV inputs for downstream BrainAGE analyses.

**Table 1 Quantitative model performance comparison across DeepContrast model architectures**

| Comparison pair/DeepCBV method | PSNR (dB) | SSIM |
|---|---|---|
| rCBV versus T1w | 10.13 | 0.778 |
| rCBV versus DeepCBV (2D RA-UNet[43]) | 29.64 | 0.879 |
| rCBV versus DeepCBV (3D RA-UNet[48]) | 29.04 | 0.842 |
| rCBV versus DeepCBV (2D TransUNet[48]) | 29.46 | 0.875 |
| rCBV versus DeepCBV (Patch-based 3D TransUNet[48]) | **30.58** | **0.912** |

The comparison includes two-dimensional (2D) and 3D RA-UNet and TransUNet variants. Performance is reported using peak signal-to-noise ratio [PSNR, decibels (dB)] and structural similarity index measure (SSIM). Higher PSNR and SSIM indicate better agreement with the ground-truth rCBV maps. The rCBV versus T1w MRI comparison provides a baseline for comparison of metrics. The patch-based 3D TransUNet achieved the highest performance (PSNR = 30.58 dB; SSIM = 0.912). Bold values indicate best performance across all models.

## Overall performance

We compared the results on the test dataset of four linear regression models created from our model architecture: (1) T1w-predicted age only (T-model), (2) DeepCBV-predicted age only (D-model), (3) combined T1w- and DeepCBV-predicted ages (TD-model) and (4) combined T1w- and DeepCBV-predicted ages with sex as an additional covariate (TDS-model), as shown in Table 2.

Table 2 contains the full results of each model created from the T1w-predicted age, DeepCBV-predicted age, and sex features with MAE, Mean Squared Error (MSE) and $R^2$. The T-model yielded a mean absolute error (MAE) of 4.10 years and $R^2$ of 0.936. The D-model (DeepCBV only) achieved an MAE of 4.49 years and $R^2$ of 0.927. Integrating both modalities in a linear regression model (combined TD-model) improved performance over the T-model and D-model, achieving an MAE of 3.96 and $R^2$ of 0.943. Adding sex as a feature in the TDS-model marginally improved performance, reducing the MAE to 3.95 and keeping the same $R^2$ of 0.943.

In Figure 4, residual plots reveal a mild age-dependent bias, with predictions tending towards the mean age at the extremes of the age range, a pattern commonly observed in brain age estimation models. This behaviour was consistent across all models and therefore does not affect relative performance comparisons. No explicit age-bias correction was applied in this analysis.

To assess the practical significance of performance differences between models, we performed nonparametric bootstrap resampling of the test set 1000 times to estimate uncertainty in RSS and differences in RSS between models. These results are also in Table 2. In all bootstrap comparisons, the 95% confidence intervals for RSS differences excluded zero, indicating statistically significant improvements for multimodal models.

The combined T1w + DeepCBV model (TD-model) showed a substantial reduction in RSS compared with the T1w-only model (RSS difference 95% CI = [349, 1719]; bootstrap $P = 0.001$), corresponding to a large effect size (Cohen's d (d) = 2.63, 95% CI: 0.98 to 4.80). Similarly, the combined model markedly outperformed the DeepCBV-only model (RSS difference = 1123 to 3081, 95% CI; $P = 0.001$), with a very large standardized effect (d = 4.02, 95% CI: 2.20–6.05).

Adding sex as a covariate to the combined model resulted in a smaller but measurable further reduction in RSS (RSS difference = 1 to 389, 95% CI; $P = 0.001$), corresponding to a moderate-to-large effect size (d = 1.19, 95% CI: 0.01 to 3.70). Together, these results indicate that while point-estimate improvements in MAE were numerically modest, the underlying reduction in prediction error was consistent, statistically robust and associated with large effect sizes.

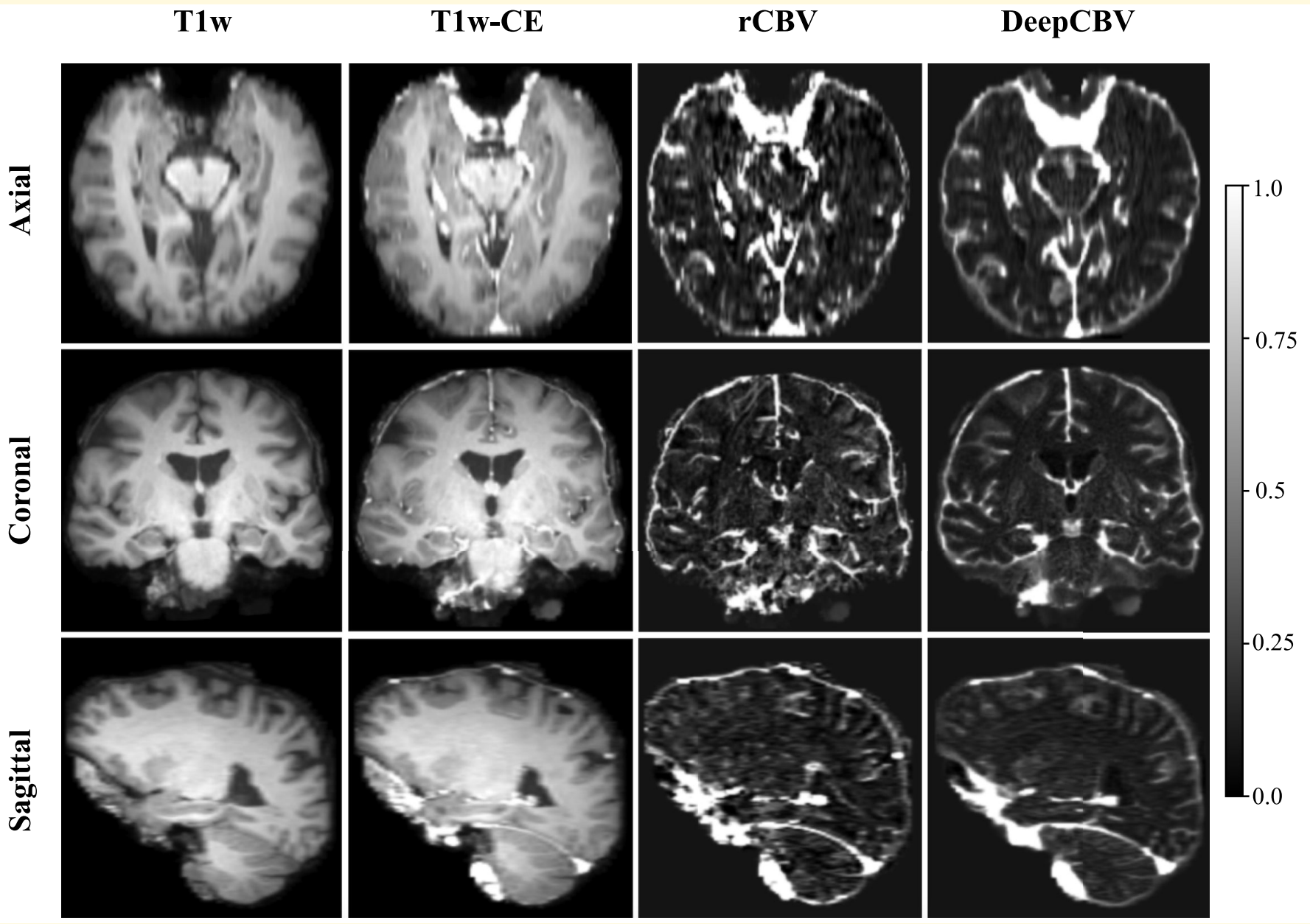

**Figure 3 DeepCBV synthesis performance evaluation.** Representative axial, coronal and sagittal slices demonstrating the DeepCBV synthesis pipeline. Columns show (left to right): T1-weighted (T1w) magnetic resonance imaging (MRI) input, contrast-enhanced T1-weighted reference (T1w-CE), relative cerebral blood volume (rCBV) map used for supervision and the predicted deep learning-derived cerebral blood volume map (DeepCBV) generated by the best-performing model (patch-based three-dimensional [3D] TransUNet). Intensity values are displayed in arbitrary units (a.u.) with the corresponding colour scale.

**Table 2 Results by model**

| Model | Mean absolute error (MAE) | Mean squared error (MSE) | $R^2$ | Comparison | RSS difference 95% CI | Effect size of RSS difference (Cohen's d) |
|---|---|---|---|---|---|---|
| T1w only (T-model) | 4.10 | 29.02 | 0.936 | T versus TD | [349.02, 1718.76][a] | 2.63 |
| DeepCBV only (D-model) | 4.49 | 33.34 | 0.927 | D versus TD | [1122.83, 3081.36][a] | 4.02 |
| T1w + DeepCBV (TD-model) | 3.96 | 26.07 | 0.943 | TD versus TDS | [1.13, 388.66][a] | 1.19 |
| T1w + DeepCBV + sex (TDS-model) | **3.95** | **25.98** | **0.943** | | | |

Performance metrics include mean absolute error (MAE), mean squared error (MSE) and R2. Model improvements were assessed using bootstrap analyses of the Residual Sum of Squares (RSS). RSS differences are reported with 95% confidence intervals (CI). [a]CI that exclude zero with statistically significant improvement ($P < 0.05$). Cohen's d reflects the standardized magnitude of RSS reduction for each model comparison. Bold values indicate best performance across all models.

## Visualizing performance

To enhance clinical interpretability, we applied Grad-CAM to visualize the regions of T1w MRI and DeepCBV inputs that most influenced the model's predictions. For each modality, we extracted the top 20% of gradient values from the final convolutional layer and overlaid them on average T1w scans across 10-year age bins.

As shown in Fig. 5, the T1w and DeepCBV encoders focus on distinct, complementary brain regions. For each modality and age group, slices were selected at the location of maximal Grad-CAM activation to highlight the brain regions most influential for the model's predictions. Slice locations vary across panels. A 3D Grad-CAM Visualization of top 20% gradient values averaged for CN participants from the test set of ages 65 to 100 from the MRI and DeepCBV CNNs to qualitatively characterize the different regions each modality emphasizes is located in the Supplementary Material. The T1w encoder emphasizes white matter structures, while the DeepCBV encoder highlights vascular-rich areas, particularly in the central and lower brain.

## Applicability to AD

To validate the clinical relevance of our BrainAGE models, we performed a series of analyses on 1233 subjects from the Alzheimer's Disease Neuroimaging Initiative (ADNI) cohort, focusing on the models' ability to capture AD-related pathology.

First, we investigated the association between BrainAGE and the diagnostic status of AD. As shown in Fig. 6, a stepwise increase in BrainAGE was observed with advancing disease severity across all models. Across diagnostic groups (CN, MCI, AD), Kruskal–Wallis tests demonstrated significant group differences for all models (T-model: H = 30.18, $P$ = 2.80 × 10 to 7; D-model: H = 33.62, $P$ = 5.00 × 10 to 8; TDS-model: H = 40.89, $P$ = 1.33 × 10 to 9), supporting robust separation across disease stages.

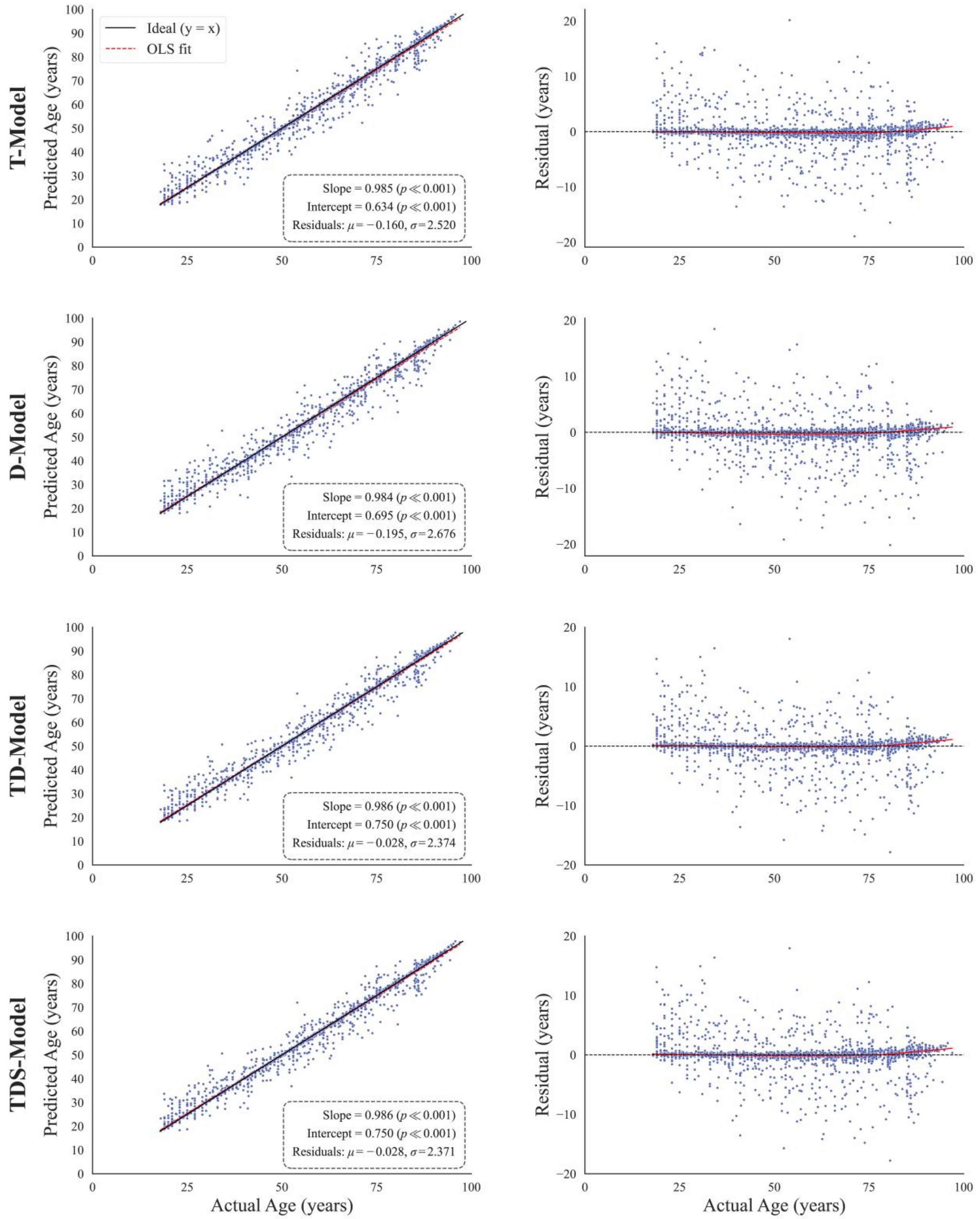

**Figure 4 Model performance for brain age prediction across modalities.** Predicted age versus chronological age (left column) and corresponding residual plots (right column) for the T-model (T1-weighted MRI), D-model (deep learning-derived cerebral blood volume; DeepCBV), TD-model (combined T1w + DeepCBV), and TDS-model (combined T1w + DeepCBV + sex). In the scatter plots, the solid line represents the identity line ($y = x$), and the dashed line represents the fitted ordinary least squares (OLS) regression line. The slope and intercept of each fitted line as well as the mean ($\mu$) and standard deviation ($\sigma$) of the fitted residuals are denoted for each subplot. All slopes and intercepts were deemed significant with $\alpha < 0.05$. Residual plots display prediction error (predicted age to chronological age) with a locally weighted smoothing curve and a horizontal dashed reference line at zero.

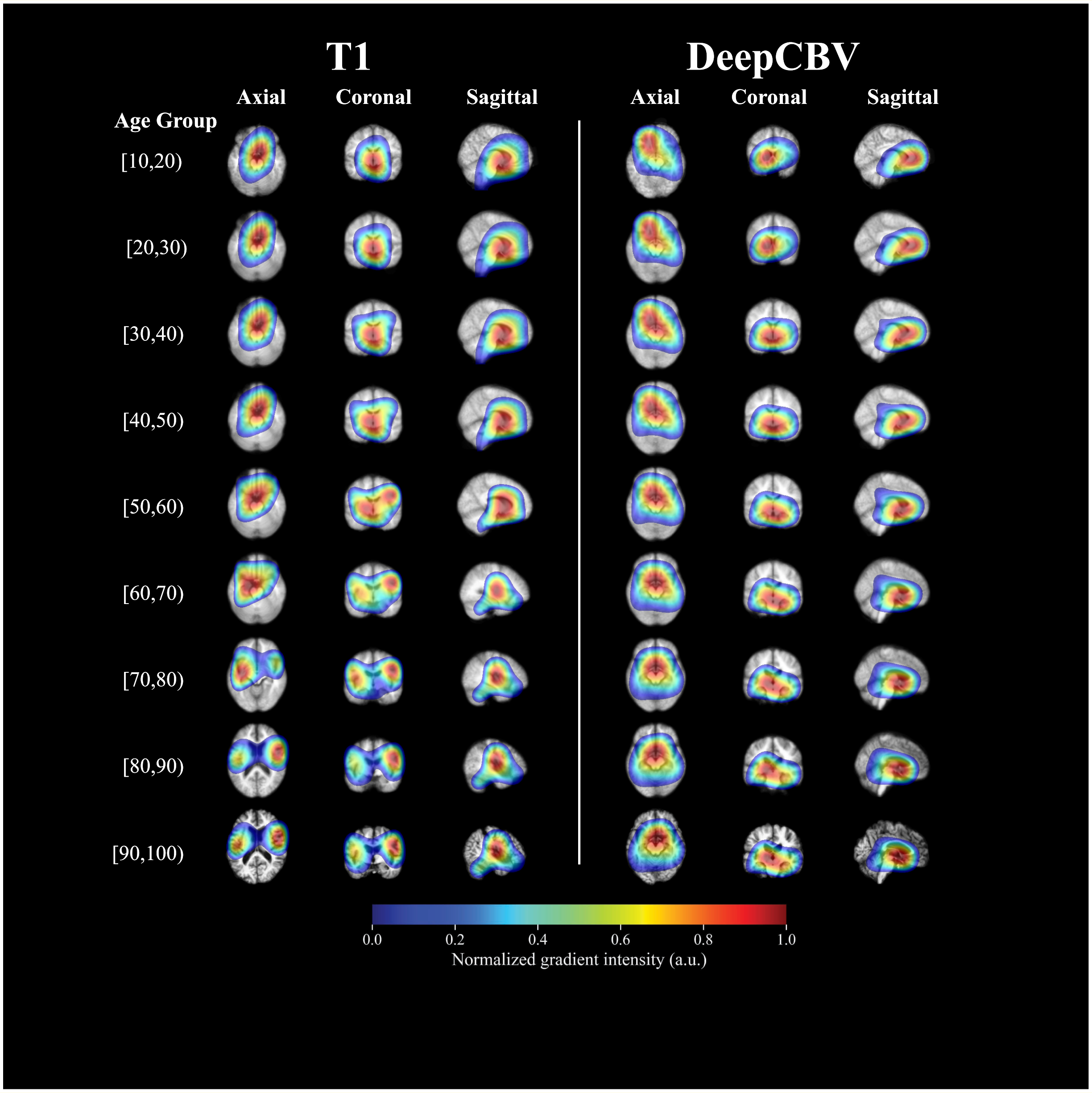

**Figure 5 Modality-specific saliency maps across age groups.** Average gradient-based saliency maps for the T1-weighted (T1w) and DeepCBV encoders across 10-year age intervals. Saliency maps were generated using gradient-weighted class activation mapping (Grad-CAM) applied to the final convolutional layer of each three-dimensional (3D) encoder. For each modality, the top 20% of voxel-wise gradient magnitudes were extracted and overlaid on the corresponding average T1w magnetic resonance imaging (MRI) scan within each age bin. For each modality and age group, slices were selected at the location of maximal Grad-CAM activation to highlight the brain regions most influential for the model's predictions and presented in radiological view. Higher gradient intensities indicate greater contribution of that region to the predicted brain age. DeepCBV denotes deep learning-derived cerebral blood volume maps synthesized from T1w MRI. Age groups are shown in 10-year intervals.

CN individuals displayed the lowest BrainAGE, followed by those with MCI, and finally, patients with AD, who exhibited the highest BrainAGE. *Post hoc* Mann–Whitney tests with Bonferroni correction demonstrated significant differences between CN and MCI for the T-model ($U = 8.037 \times 10^4$, $P = 7.73 \times 10^{-4}$), D-model ($U = 8.439 \times 10^4$, $P = 3.57 \times 10^{-2}$), and TDS-model ($U = 8.004 \times 10^4$, $P = 5.37 \times 10^{-4}$). We also report the effect size calculated using Cohen's d (d) as 0.27, 0.20 and 0.28 for the T-model, D-model and TDS-model respectively. Comparisons between CN and AD were highly significant across all models, including the T-model ($d = 0.49$, $U = 2.128 \times 10^4$, $P = 1.55 \times 10^{-6}$), D-model ($d = 0.55$, $U = 2.026 \times 10^4$, $P = 3.68 \times 10^{-8}$), and TDS-model ($d = 0.58$, $U = 1.965 \times 10^4$, $P = 3.34 \times 10^{-9}$).

For MCI versus AD, significant differences were observed for the D-model ($d = 0.33$, $U = 4.846 \times 10^4$, $P = 7.02 \times 10^{-4}$) and TDS-model ($d = 0.30$, $U = 4.913 \times 10^4$, $P = 1.82 \times 10^{-3}$), whereas the T-model did not show a statistically significant difference between these groups ($d = 0.23$, $U = 5.21 \times 10^4$, $P = 0.0661$). Overall, the TDS-model demonstrated the strongest and most consistent statistical separation across diagnostic strata.

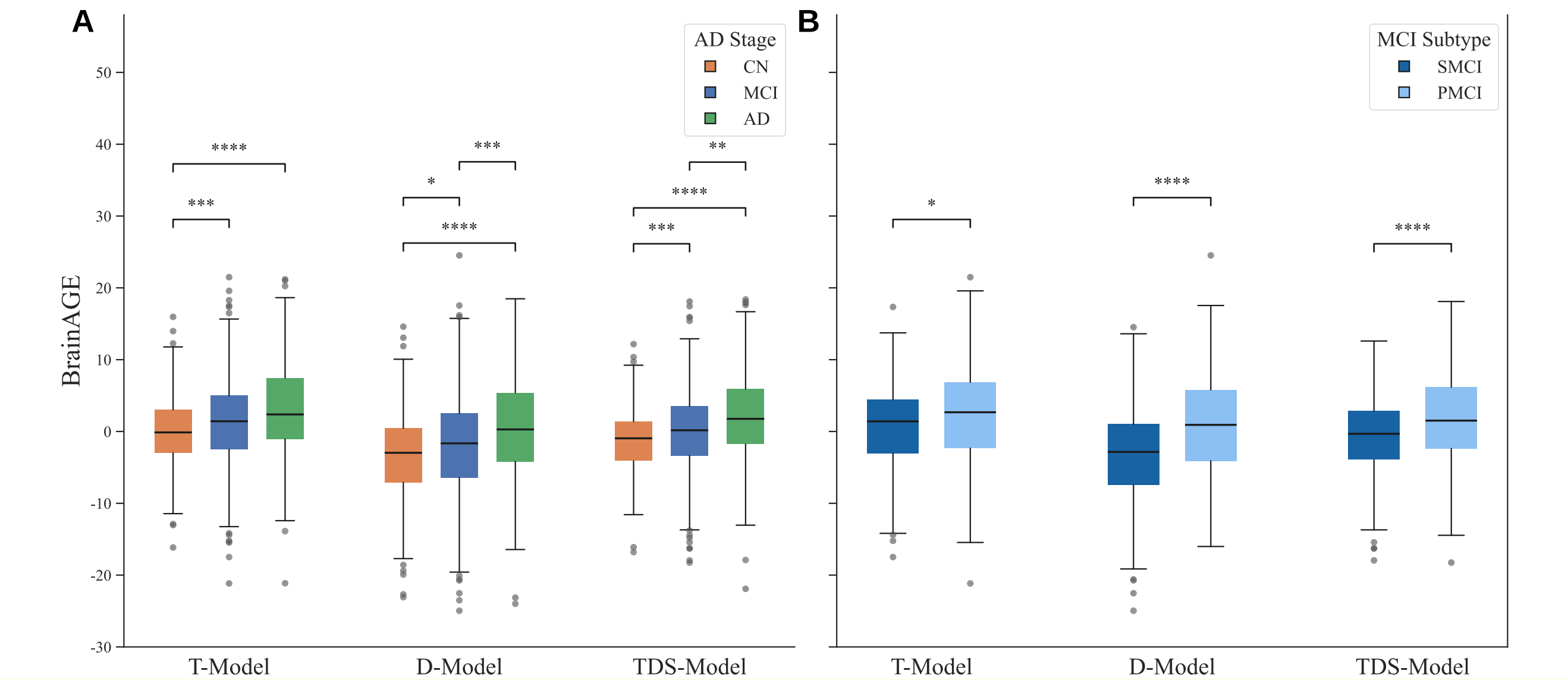

**Figure 6 Boxplot for BrainAGE scores by Alzheimer's disease risk.** A boxplot showing the distribution of BrainAGE scores for the T-model, D-model, and TDS-model across (**A**) cognitively normal (CN), mild cognitive impairment (MCI) and Alzheimer's Disease (AD) and (**B**) stable MCI (SMCI) and progressive MCI (PMCI) with respective post-hoc Mann–Whitney tests with Bonferroni correction to show statistical significance. Statistical significance is indicated as follows: $P < 0.05$ (*), $P < 0.01$ (**), $P < 0.001$ (***) and $P < 0.0001$ (****).

Next, we evaluated the prognostic potential of the models to differentiate between individuals with MCI who remained stable (SMCI) versus those who progressed to AD (PMCI). Focusing specifically on MCI progression, Mann–Whitney U tests revealed that baseline BrainAGE was significantly higher in the PMCI group compared to the SMCI group across all three models. The T-model showed a significant difference ($d = 0.30$, $U = 2.64 \times 10^4$, $P = 1.26 \times 10^{-2}$). The D-model demonstrated the strongest effect ($d = 0.54$, $U = 2.18 \times 10^4$, $P = 4.43 \times 10^{-8}$), followed by the TDS-model ($d = 0.46$, $U = 2.33 \times 10^4$, $P = 7.25 \times 10^{-6}$), highlighting the added value of incorporating DeepCBV-derived information for predicting disease progression.

Finally, we assessed the relationship between BrainAGE and established clinical scores using Spearman correlation analysis (Table 3). The BrainAGE values demonstrated significant associations with measures of dementia severity and cognitive function. BrainAGE was positively correlated with the Clinical Dementia Rating Sum of Boxes (CDRSB), indicating that an older-appearing brain corresponds to greater functional impairment. This relationship was significant for the T-model ($\rho = 0.477$, $P < 0.001$) and TDS-model ($\rho = 0.403$, $P = 0.005$) models. Conversely, BrainAGE was negatively correlated with the Mini-Mental State Examination (MMSE) score, where a higher BrainAGE is associated with poorer cognitive performance. This was significant for the DeepCBV BrainAGE ($\rho = -0.369$, $P = 0.008$) and Combo BrainAGE ($\rho = -0.310$, $P = 0.029$) models. No significant correlations were observed for the ADAS13 scores.

**Table 3 Cognitive tests and model associations**

| Variable | BrainAGE model | Spearman ρ | *P*-value |
|---|---|---|---|
| CDRSB | T-Model | 0.477* | 0.000609* |
| | D-Model | 0.241 | 0.0995 |
| | TDS-Model | 0.403* | 0.00456* |
| MMSE | T-Model | −0.186 | 0.196 |
| | D-Model | −0.369* | 0.00828* |
| | TDS-Model | −0.310* | 0.0286* |
| ADAS13 | T-Model | 0.175 | 0.225 |
| | D-Model | 0.215 | 0.134 |
| | TDS-Model | 0.190 | 0.187 |

The Spearman correlation coefficients ρ and associated *P*-values for the relationship between BrainAGE values and various cognitive scores, including the Clinical Dementia Rating Sum of Boxes (CDRSB), the Mini-Mental State Examination (MMSE) and the Alzheimer's Disease Assessment Scale (ADAS).

# Discussion

In this study, we developed and validated a multimodal deep learning framework for BrainAGE that synergistically combines structural T1w MRI with DeepCBV maps. The results offer profound insights into the distinct and complementary biological processes that characterize brain ageing. The primary contribution of this study is the demonstration that a learned vascular representation derived from non-contrast T1-weighted MRI can be used to generate biologically meaningful BrainAGE estimates. DeepCBV-based BrainAGE captures age-related signal and provides strong discrimination between stable and progressive mild cognitive impairment, supporting its relevance for modelling disease-related ageing processes. By integrating DeepCBV with structural MRI, this framework highlights the value of vascular information for BrainAGE and clinical AD risk stratification. This joint modelling of complementary imaging information more closely captures brain age and pathological effects on brain age, which allows the output of the TDS-model to serve as a biomarker for early detection of AD progression. The model assigned a small but statistically significant negative coefficient ($\beta = -0.168$, $P = 0.02$) to the male sex, suggesting that male brains are consistently predicted as being biologically older than female brains of the same chronological age. This result aligns with established clinical evidence of sex-differentiated brain ageing trajectories, further grounding our model's predictions in known physiological patterns.[55-57]

The rationale for this multimodal approach is grounded in the limitations of unimodal imaging. T1w MRI is effective at capturing macroscopic anatomical changes like cortical and subcortical atrophy, which are well-established hallmarks of brain ageing.[58] However, these structural alterations are often late-stage manifestations of underlying ageing-related signals.[59,60] The subtle physiological processes that precede them, such as changes in vasculature, remain largely invisible to conventional structural scans.[60-62] The DeepCBV maps directly address this gap by providing a proxy for these functional dynamics. Our framework generates this vital functional information from standard, non-contrast T1w scans, thereby circumventing the significant clinical and logistical barriers associated with traditional functional imaging techniques. By leveraging a deep learning model to synthesize CBV maps, our approach democratizes access to functional brain information, making a multimodal assessment scalable, safe for repeated measures and applicable to open-source neuroimaging databases.

To understand the mechanistic basis for this enhanced performance, we employed gradient-based class activation maps (Grad-CAM) to visualize the brain regions that most influenced each model's predictions. The T1w encoder consistently focused its attention on white matter structures, learning to detect age-related changes in white matter integrity and cortical loss.[63] In contrast, the DeepCBV encoder preferentially highlighted vascular-rich areas, particularly in the periventricular areas, capturing information about blood flow and metabolic activity. This confirms that the two modalities used in the TDS-model provide unique, non-redundant information, explaining why the combined model outperforms its unimodal counterparts.

The Grad-CAM analysis uncovered a biologically plausible dynamic shift in regional focus across the human lifespan, a finding that aligns with prominent neurocognitive theories of ageing.[64] For the T1w-only model, the predictive focus in younger individuals was predominantly in the medial temporal lobe. As age increased, however, the model's attention transitioned to the inferior frontal cortices (IFC). The increased reliance on the IFC in older age suggests the model has learned that this region becomes a more salient predictor of age, perhaps following patterns of atrophy concordant with ageing theory.[58]

A parallel and compelling age-related transition was observed for the DeepCBV model. In younger populations, the focus was on the prefrontal cortex, likely reflecting the high metabolic demands of executive functions. In older adults, however, the model's predictive focus shifted to deep brain structures, particularly the medial temporal region, indicative of the model learning to detect key patterns of age-related cerebrovascular decline.[60] This resonates with theories of neurocognitive ageing, namely Posterior-to-Anterior Shift in Ageing (PASA), which is a functional model describing how older adults compensate for posterior sensory deficits by recruiting anterior prefrontal regions. In fact, while adhering to PASA,[65] the DeepCBV model's shift in focus also follows patterns of vascular damage in periventricular regions where vasculature vulnerability increases with age. Further, the periventricular region is a classic location for white matter hyperintensities (WMH), which are lesions indicative of chronic ischaemia that are associated with more advanced age and cognitive decline.[6-8] The enlargement of the lateral ventricles with age is a well-known hallmark of ageing,[58] and the model's increased focus on the adjacent tissue suggests a particular sensitivity to vascular damage. Specifically, the DeepCBV model learned to identify regions of compromised vasculature and likely hypoperfusion as powerful predictors of brain ageing. This validates that the DeepCBV-based predictions capture distinct and clinically relevant markers of cerebrovascular health, providing functional insights that complement the purely structural features identified by the T1w model.

Our combined TDS-model appears to offer a heightened sensitivity to the earliest signs of neurodegeneration. In diseases like AD, reduced cerebral blood flow and brain atrophy both contribute to the onset of neurodegeneration.[66-68] The central aim of this study was to move beyond creating a statistically accurate age prediction model and develop a tool with tangible clinical relevance for age-related neurodegenerative diseases. By applying our models to a large, well-characterized cohort from ADNI, we have demonstrated that integrating DeepCBV with traditional T1w structural MRI provides a substantially more powerful biomarker for AD than either modality alone. The clear increase in BrainAGE from CN individuals to those with MCI and finally to patients with diagnosed AD aligns with the conceptualization of AD as a continuum of advancing neuropathology.

The T1w-based BrainAGE correlated most strongly with the CDRSB, which evaluates dementia severity, while the DeepCBV-based BrainAGE correlated most strongly with MMSE, a direct measure of cognitive status. This suggests that the vascular information captured by DeepCBV may be more tightly linked to the subtle network disruptions that manifest as cognitive deficits, particularly in the earlier stages of the disease, while dementia progression may be tightly coupled to structural changes. However, only the combined BrainAGE model was able to demonstrate significant correlations with both the CDRSB and the MMSE, which indicates that a multimodal approach successfully integrates the strengths of both models, presenting a biomarker that potentially reflects cognitive impairment.

A potentially significant clinical contribution of this work is the model's enhanced ability to predict which individuals with MCI progress to AD. Our results show that while the T-model could significantly differentiate between SMCI and PMCI, its predictive power was dwarfed by the models incorporating DeepCBV. The D-model was the single strongest predictor by a substantial margin, with the combined model also showing improvement over the T-model as a result. This finding strongly suggests that at the MCI stage, functional and vascular changes are more sensitive markers of impending decline than the structural changes alone. However, the combined TDS-model demonstrated improved risk stratification in AD, integrating the strengths of both single modality models, which may allow for the enrichment of clinical trials with high-risk individuals and enable earlier, more targeted patient management.

Although the absolute improvements in MAE between models were modest and do not exceed current benchmarks,[19] prior work has shown that brain-age models optimized exclusively for minimal age prediction error may exhibit reduced sensitivity to disease-related effects, whereas models with modestly lower age prediction accuracy can show enhanced sensitivity for disease detection and group discrimination.[69] Importantly, bootstrap analyses revealed that these small gains in MAE still corresponded to large and consistent reductions in residual error with effect sizes also indicating substantial improvements in model performance. From a clinical perspective, these observed improvements reflect not mere incremental optimization of age prediction accuracy, but increased reliability and clinical utility of BrainAGE for AD risk stratification.

A key limitation of this study is the use of generated vascular data. DeepCBV maps are synthesized from non-contrast T1-weighted MRI and therefore represent a learned vascular proxy rather than an independently acquired perfusion measurement. DeepCBV is derived from structural MRI to recreate contrast-enhanced data, therefore may still partially re-encode anatomical information already present in the T1-weighted image rather than capturing purely physiological vascular signal. Accordingly, DeepCBV-based BrainAGE should be interpreted as reflecting vascularly informed patterns of brain ageing rather than direct cerebral blood volume measurements. Although our analyses demonstrated that DeepCBV contributes clinically relevant vascular information, future work incorporating independently acquired perfusion imaging may be necessary to further characterize the physiological efficacy and specificity of generated vascular maps.

In this study, we demonstrated that integrating anatomical (T1w MRI) and functional (DeepCBV) data improves BrainAGE prediction accuracy, validating the synergistic value of multimodal fusion of structural and functional imaging data. Future studies will further validate DeepCBV as a clinically deployable surrogate for CBV-fMRI, enabling non-contrast assessment of regional basal brain activity and metabolic demand. Combining robust modelling strategies with translational validation will help bridge the gap between research and clinical practice, advancing BrainAGE as a clinically actionable biomarker.

# Supplementary material

Supplementary material is available at *Brain Communications* online.

# Funding

This research was conducted while Jia Guo was a Hevolution Foundation/American Federation for Aging Research New Investigator Awardee in Aging Biology and Geroscience Research [PT# 24-4274 PG014621]. This work was supported by a grant from the Simons Foundation [811245, S.A.S.] and a grant from the Simons Foundation [SFI-AN-NC-AB-Research-00018029, S.A.S.].

# Competing interests

The authors report no competing interests.

# Data availability

Data were provided in part by IXI, accessed from http://brain-development.org/ixi-dataset.

Data were provided in part by OASIS. OASIS Cross-Sectional: Principal Investigators: D. Marcus, R, Buckner, J, Csernansky J. Morris; P50 AG05681, P01 AG03991, P01 AG026276, R01 AG021910, P20 MH071616, U24 RR021382 OASIS: Longitudinal: Principal Investigators: D. Marcus, R, Buckner, J. Csernansky, J. Morris; P50 AG05681, P01 AG03991, P01 AG026276, R01 AG021910, P20 MH071616, U24 RR021382.

Data were provided in part by the Brain Genomics Superstruct Project of Harvard University and the Massachusetts General Hospital (Principal Investigators: Randy Buckner, Joshua Roffman, and Jordan Smoller), with support from the Center for Brain Science Neuroinformatics Research Group, the Athinoula A. Martinos Center for Biomedical Imaging, and the Center for Human Genetic Research. 20 individual investigators at Harvard and MGH generously contributed data to the overall project.

Data used in preparation of this article were obtained from the Alzheimer's Disease Neuroimaging Initiative (ADNI) database (adni.loni.usc.edu). As such, the investigators within the ADNI contributed to the design and implementation of ADNI and/or provided data but did not participate in analysis or writing of this report. A complete listing of ADNI investigators can be found at: http://adni.loni.usc.edu/wp-content/uploads/how_to_apply/ADNI_Acknowledgement_List.pdf

Data collection and sharing for this project was funded by the Alzheimer's Disease Neuroimaging Initiative (ADNI) (National Institutes of Health Grant U01 AG024904) and DOD ADNI (Department of Defense award number W81XWH-12–2-0012). ADNI is funded by the National Institute on Aging, the National Institute of Biomedical Imaging and Bioengineering, and through generous contributions from the following: AbbVie, Alzheimer's Association; Alzheimer's Drug Discovery Foundation; Araclon Biotech; BioClinica, Inc.; Biogen; Bristol-Myers Squibb Company; CereSpir, Inc.; Cogstate; Eisai Inc.; Elan Pharmaceuticals, Inc.; Eli Lilly and Company; EuroImmun; F. Hoffmann-La Roche Ltd and its affiliated company Genentech, Inc.; Fujirebio; GE Healthcare; IXICO Ltd.; Janssen Alzheimer Immunotherapy Research & Development, LLC.; Johnson & Johnson Pharmaceutical Research & Development LLC.; Lumosity; Lundbeck; Merck & Co., Inc.; Meso Scale Diagnostics, LLC.; NeuroRx Research; Neurotrack Technologies; Novartis Pharmaceuticals Corporation; Pfizer Inc.; Piramal Imaging; Servier; Takeda Pharmaceutical Company; and Transition Therapeutics. The Canadian Institutes of Health Research is providing funds to support ADNI clinical sites in Canada. Private sector contributions are facilitated by the Foundation for the National Institutes of Health (www.fnih.org). The grantee organization is the Northern California Institute for Research and Education, and the study is coordinated by the Alzheimer's Therapeutic Research Institute at the University of Southern California. ADNI data are disseminated by the Laboratory for Neuro Imaging at the University of Southern California.

Data used in the preparation of this article was obtained from the Australian Imaging Biomarkers and Lifestyle flagship study of ageing (AIBL) funded by the Commonwealth Scientific and Industrial Research Organisation (CSIRO) which was made available at the ADNI database (adni.loni.usc.edu). The AIBL researchers contributed data but did not participate in analysis or writing of this report. AIBL researchers are listed at www.aibl.csiro.au.

Data used in preparation of this article were obtained from the Frontotemporal Lobar Degeneration Neuroimaging Initiative (FTLDNI) database (https://ida.loni.usc.edu/login.jsp?project=NIFD). The investigators at NIFD/FTLDNI contributed to the design and implementation of FTLDNI and/or provided data, but did not participate in analysis or writing of this report (unless otherwise listed).

Data collection and sharing for this project was funded by the Frontotemporal Lobar Degeneration Neuroimaging Initiative (National Institutes of Health Grant R01 AG032306). The study is coordinated through the University of California, San Francisco, Memory and Aging Center. FTLDNI data are disseminated by the Laboratory for Neuro Imaging at the University of Southern California.

Data used in the preparation of this article were obtained from the Parkinson's Progression Markers Initiative (PPMI) database (https://www.ppmi-info.org/access-data-specimens/data). For up-to-date information on the study, visit www.ppmi-info.org. PPMI—a public–private partnership—is funded by the Michael J. Fox Foundation for Parkinson's Research and funding partners, the full names of all of the PPMI funding partners can found at https://www.ppmi-info.org/about-ppmi/who-we-are/study-sponsors.

Data used in preparation of this article were obtained from the SchizConnect database (http://schizconnect.org) As such, the investigators within SchizConnect contributed to the design and implementation of SchizConnect and/or provided data but did not participate in analysis or writing of this report. Data collection and sharing for this project was funded by NIMH cooperative agreement 1U01MH097435.

All data used during the study are available in the International Data-sharing Initiative (INDI, http://fcon_1000.projects.nitrc.org/) provided by the Southwest University Longitudinal Imaging Multimodal (SLIM) study.

# Appendix

Data used in preparation of this article were obtained from the Frontotemporal Lobar Degeneration Neuroimaging Initiative (FTLDNI) database. The investigators at NIFD/FTLDNI contributed to the design and implementation of FTLDNI and/or provided data, but did not participate in analysis or writing of this report (unless otherwise listed). The FTLDNI investigators included the following individuals:

Howard Rosen, Bradford C. Dickerson, Kimoko Domoto-Reilly, David Knopman, Bradley F. Boeve, Adam L. Boxer, John Kornak, Bruce L. Miller, William W. Seeley, Maria-Luisa Gorno-Tempini, Scott McGinnis and Maria Luisa Mandelli.